\documentclass{jfm}
\usepackage{graphicx}
\usepackage{epstopdf, epsfig}
\usepackage{bm}
\usepackage{amsmath,amsfonts}
\usepackage{hyperref} 
\usepackage{caption}
\usepackage{subcaption}
\usepackage{csquotes}
\usepackage{xcolor}

\title{Unsteady motion past a sphere translating steadily in wormlike micellar solutions: A numerical analysis}

\author{Chandi Sasmal\aff{1}
  \corresp{\email{csasmal@iitrpr.ac.in}},
  }

\affiliation{\aff{1}Soft Matter Engineering and Microfluidics Lab, Department of Chemical Engineering, Indian Institute of Technology
Ropar, Rupnagar, India-140001}

\begin{document}
\maketitle
\begin{abstract}
This study numerically investigates the flow characteristics past a solid and smooth sphere translating steadily along the axis of a cylindrical tube filled with wormlike micellar solutions in the creeping flow regime. The two-species VCM (Vasquez-Cook-McKinley) and single-species Giesekus constitutive models are used to characterize the rheological behaviour of micellar solutions. Once the Weissenberg number exceeds a critical value, an unsteady motion downstream of the sphere is observed in the case of two-species model. We provide the evidence that this unsteady motion downstream of the sphere is caused due to the sudden rupture of long and stretched micelles in this region, resulting from an increase in the extensional flow strength. The corresponding single-species Giesekus model for the wormlike micellar solution, with no breakage and reformation, predicts a steady flow field under otherwise identical conditions. Therefore, it further ascertains the evidence presented herein for the onset of this unsteady motion. Furthermore, we find that the onset of this unsteady motion downstream of the sphere is delayed as the sphere to tube diameter ratio decreases. A similar kind of unsteady motion has also been observed in earlier several experiments for the problem of a sphere sedimenting in a tube filled wormlike micellar solutions. We find a remarkable qualitative similarity in the flow characteristics between the present numerical results on the steadily translating sphere and prior experimental results on the falling sphere.       
\end{abstract}


\section{\label{Intro}Introduction}
A solid sphere translating in a cylindrical tube filled with a quiescent liquid is one of the classical and benchmark problems in the helm of transport phenomena for many decades. It represents an idealization of many industrially relevant processes, for instance, fluidized and fixed bed reactors, slurry reactors, falling ball viscometer, equipment for separating solid-liquid mixture in mining, and petrochemical industries, processing of polymer melts, etc. Not only of practical significance, but this problem is also of fundamental interest in its own right. As a result, this problem has been extensively investigated in the research community, and much has been written about it in the literature for both Newtonian and non-Newtonian fluids~\citep{mckinley2002steady,chhabra2006bubbles,michaelides2006particles}. Earlier investigations on this problem were restricted to simple Newtonian fluids like water, and it has been then gradually extended to complex non-Newtonian fluids like polymer solutions and melts due to their overwhelming applications in scores of industrial settings like food, petrochemical, personal care products, etc~\citep{chhabra2006bubbles}. Earlier investigations revealed that both the blockage ratio (ratio of the sphere to tube diameter) and non-linear rheological properties of fluids like shear-thinning, shear-thickening, visco-plasticity, etc. greatly influenced the flow characteristics like the drag force, wake length, etc., in comparison to that for an unconfined situation and for Newtonian fluids. In addition to the investigations carried out for the generalized Newtonian fluids, many studies have also been presented for viscoelastic fluids. Some typical and complex flow features were seen in these fluids as compared to that seen either in Newtonian or any GNF fluid. This complexity was not only observed in the variation of the integral parameters like the drag force but also seen in the flow fields near the sphere. For instance, a downward or upward shifting in the axial velocity profile along the upstream or downstream axis of the sphere has been observed both experimentally and numerically for viscoelastic fluids in comparison to that seen in Newtonian fluids~\citep{arigo1995sedimentation,arigo1998experimental,bush1994stagnation}. Additionally, a flow reversal phenomenon and/or the presence of a “negative wake” downstream of the sphere has also been observed in viscoelastic fluids~\citep{harlen2002negative,bisgaard1983velocity}.

The next generation of studies on this benchmark problem has considered a solution comprised of various types of surfactant molecules. When these molecules dissolve in water above a critical concentration, they spontaneously self-assemble into large and flexible aggregates of micelles of different shapes like spherical, ellipsoidal, wormlike, or lamellae~\citep{moroi1992micelles}. The rheological properties of these wormlike micellar (WLM) solutions  were found to be more complex than that seen for polymer solutions or melts~\citep{rothstein2008strong,rothstein2003transient}. This is due to the fact that these wormlike micelles can undergo continuous scission and reformation in an imposed shear or extensional flow field, unlike polymer molecules, which are unlikely to break due to the presence of a strong covalent backbone. Because of their extensive applications over a wide range of industrial settings, a considerable amount of studies have also been performed on the falling sphere problem in these fluids in the creeping flow regime. For instance, Jayaraman and Belmonte~\citep{jayaraman2003oscillations} conducted an experimental investigation on this problem in CTAB (Cetyl trimethyl ammonium bromide)/NaSal(Sodium salicylate) wormlike micellar solution. They found an unsteady motion of the sphere in the direction of its sedimentation. They proposed that the cause for this instability was due to the destruction of the flow-induced structure (FIS) formed in the sphere's vicinity. However, in a later study with the same WLM solution, Chen and Rothstein ~\citep{chen2004flow} claimed that this instability was due to the sudden rupture of the long micelles downstream of the sphere. This reason was further established in a later study with CPyCl (Cetylpyridinium chloride)/NaSal WLM solution by  Wu and Mohammadigoushki~\citep{wu2018sphere}. The unsteady motion of the falling sphere has also been observed in the study by Kumar et al.~\citep{kumar2012} with CTAT (Cetyl trimethyl ammonium p-toluenesulphonate)/NaCl (Sodium chloride) micellar solution and in a recent study by Wang et al.~\citep{wang2020} with  OTAC (Octadecyl trimethyl ammonium chloride)/NaSal micellar solution. To characterize the falling sphere's onset of this unsteady motion, Mohammadigoushki and Muller~\citep{mohammadigoushki2016sedimentation} and Zhang and Muller~\citep{zhang2018unsteady} found out a criteria by  calculating the local extensional Weissenberg number downstream of the sphere based on the local maximum extension rate. This criterion is found to be universally valid as they discovered that it doesn't depend on the micelles chemistry and solution rheological behaviours, for instance, whether the solution shows shear banding phenomena or not. Furthermore, their predictions for the unsteady motion were in line with that predicted by Chen and Rothstein~\citep{chen2004flow} and Wu and Mohammadigoushki~\citep{wu2018sphere}.

Therefore, most of the studies proposed that the unsteady motion of a sedimenting sphere in wormlike micellar solutions is due to the presence of strong extensional flow downstream of the sphere, causing the sudden rupture of highly aligned and stretched micelles in this region~\citep{rothstein2008strong}. However, there is no direct evidence present on this, and it was only indirectly proved using the FIB (flow induced birefringence) and PIV (particle image velocimetry) experiments~\citep{chen2004flow,wu2018sphere}. The present study aims to establish this hypothesis using numerical simulations based on the Vasquez-Cook-McKinley (VCM) constitutive model~\citep{vasquez2007network} for the wormlike micellar solution. However, it should be mentioned here that the problem considered in this study is not the exact representation of the prior experimental settings wherein the sphere is allowed to sediment in the tube due to its own weight~\citep{chen2004flow,wu2018sphere,zhang2018unsteady,jayaraman2003oscillations}. The sphere may rotate or undergo lateral motion during the sedimentation or even it may not reach to a terminal velocity~\citep{mohammadigoushki2016sedimentation}. Therefore, in actual experiments, the flow may become three-dimensional and non-axisymmetric. To realize the corresponding experimental conditions accurately, one has to solve numerically the full governing field equations, namely, continuity, momentum and micellar constitutive equations in a three dimensional computational domain along with an equation of the sphere motion. In the present simulations, we consider a problem wherein the sphere is translating steadily along the axis of a tube, and this can be a situation in the corresponding experiments on the falling sphere problem when the sphere will reach a terminal velocity. Although this is not a case in actual experiments; however, by using this simplified problem, we aim to show that this unsteady motion  downstream of the sphere is, indeed, caused due to the breakage of micelles. Therefore, this will further establish the hypothesis for the unsteady motion of the sphere falling in wormlike micellar solutions, as observed in prior experiments~\citep{chen2004flow,wu2018sphere,zhang2018unsteady,jayaraman2003oscillations}.    

To prove the aforementioned hypothesis, as stated above, the present study plans to use the two-species VCM constitutive model for characterizing the rheological behaviour of wormlike micellar solutions. This model considers the micelles as elastic segments composed of Hookean springs, which all together form an elastic network. The breakage and reformation dynamics were incorporated in this model based on Cate's original reversible breaking theory for wormlike micelles~\citep{cates1987reptation}. For different viscometric flows, a very good agreement has been found between the predictions obtained with the VCM model and the corresponding experimental results~\citep{pipe2010wormlike,mohammadigoushki2019transient} whereas for a non-viscometric complex flow, a good qualitative correspondence has been seen in recent studies~\citep{sasmal2020flow,khan2020effect,kalb2017role,kalb2018elastic}. Therefore, this VCM model's capability to predict the flow behaviour of wormlike micellar solution in various flow fields is well established.  We also use the single-species Giesekus constitutive equation in our analysis to show the importance of breakage and reformation of micelles for the onset of this unsteady motion.  

\section{\label{ProbFor}Problem formulation and governing equations}
\begin{figure}
    \centering
    \includegraphics[trim=0cm 0cm 0cm 0cm,clip,width=12cm]{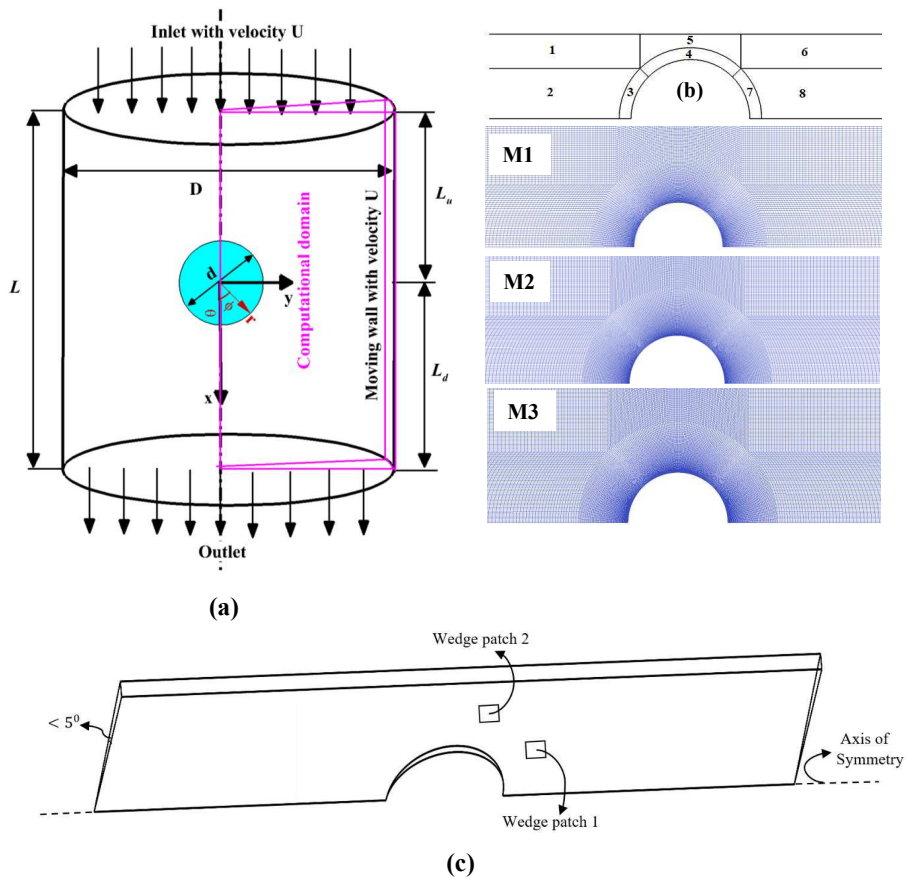}
    \caption{(a) Schematic of the present problem with both Cartesian and spherical coordinates (b) Different mesh densities used in the present study with a zoomed view near the sphere surface (c) Implication of the wedge boundary condition to approximate two-dimensional and axisymmetric condition of the present problem in OpenFOAM.}
    \label{fig:Figure1}
\end{figure}

The problem considered herein is the study of the flow characteristics of a sphere of diameter $d$, translating steadily along the axis of a cylindrical tube of diameter $D$ filled with an incompressible wormlike micellar solution in the creeping flow regime, as schematically shown in figure~\ref{fig:Figure1}(a). 
The present problem is solved in an Eulerian reference frame wherein the coordinate system is centered on and traveling with the sphere. In this coordinate system, the velocity vector is assumed to be zero on the sphere surface. Furthermore, at the inlet and tube walls, the dimensionless fluid axial velocity is set to be unity, and the radial velocity is set to be zero (discussed in detail in the subsequent section) as schematically shown in figure~\ref{fig:Figure1}(a). Furthermore, the flow is assumed to be two-dimensional and axisymmetric in nature.

Two values of the blockage ratio (the ratio of the sphere diameter to that of the tube diameter, i.e.,  $d/D$), namely, 0.33 and 0.1, are considered in this study, whereas the upstream $(L_{u})$ and downstream $(L_{d})$ lengths of the tube are chosen as 65$d$. These values are sufficiently high so that the end effects are negligible. This was further confirmed by performing a systematic domain independence study. 
\subsection{Flow equations}
Under the circumstances mentioned above, the flow field will be governed by the following equations written in their dimensionless forms as follows:\newline
Equation of continuity
\begin{equation}
\label{mass}
    \bm{\nabla} \cdot \bm{U} = 0
\end{equation}
Cauchy momentum equation
\begin{equation}
\label{mom}
   El^{-1} \frac{D\bm{U}}{Dt} = -\nabla P + \nabla \cdot \bm{\tau}
\end{equation}
In the above eqs, $\bm{U}$, $t$ and $\bm{\tau}$ are the velocity vector, time and total extra stress tensor respectively, whereas $El$ is the elasticity number defined at the end of this section. For an intertialess flow, the left hand side of equation~\ref{mom} is essentially zero. The total extra stress tensor, $\bm{\tau}$, for a wormlike micellar solution is given as:
\begin{equation}
    \label{totalStress}
    \bm{\tau} = \bm{\tau_{w}} + \bm{\tau_{s}}  
\end{equation}
where $\bm{\tau_{w}}$ is the non-Newtonian contribution from the wormlike micelles  whereas $\bm{\tau_{s}}$ is the contribution from that of the Newtonian solvent which is equal to $\beta \dot{\bm{\gamma}}$. Here the parameter $\beta$ is the ratio of the solvent viscosity to that of the zero-shear rate viscosity of the wormlike micellar solution and $\dot{\bm{\gamma}} = \nabla \bm{U} + \nabla \bm{U}^{T} $ is the strain-rate tensor. For the two-species VCM model, the total extra stress tensor is given by 
\begin{equation}
     \bm{\tau} = \bm{\tau}_{w}^{VCM} + \bm{\tau_{s}} = (\bm{A} + 2\bm{B}) - \left(n_{A} + n_{B}\right)\bm{I} + \beta_{VCM}\dot{\bm{\gamma}}
\end{equation}
Here $n_{A}$ and $\bm{A}$ are the number density and conformation tensor of the long worm A respectively, whereas $n_{B}$ and $\bm{B}$ are to that of the short worm B in the two-species model. The temporal and spatial evaluation of the number density and conformation tensor of worms are written in the following subsection. For the single-species Giesekus model, this is given by 
\begin{equation}
    \bm{\tau} = \bm{\tau}_{w}^{G} + \bm{\tau_{s}} = ( \bm{A} - \bm{I}) + \beta_{G}\dot{\bm{\gamma}}
\end{equation}
 Note that here all the lengths, velocity, time and conformation tensors are non-dimensionalized using $d$, $d/\lambda_{eff}$, $\lambda_{eff}$, and $G_{0}^{-1}$ respectively, where $\lambda_{eff} = \frac{\lambda_{A}}{1+c_{Aeq}^{'}\lambda_{A}}$ is the effective relaxation time in the two-species VCM model, $G_{0}$ is the elastic modulus, $\lambda_{A}$ and $c_{Aeq}^{'}$ are the dimensional relaxation time and equilibrium breakage rate of the long chain A respectively. In case of the single species model, $\lambda_{eff}$ is replaced by the Maxwell relaxation time $\lambda$ during the non-dimensionalization. The elasticity number is defined $El = \frac{Wi}{Re}$, where $Wi_{S} = \frac{\lambda_{eff}U}{d}$ is the shear Weissenberg number and $Re = \frac{d U \rho}{\eta_{0}}$ is the Reynolds number.
 
 \subsection{Two-species VCM constitutive equations}
 
 The VCM constitutive equations provide the species conservation equations for the long $(n_{A})$ and short worms $(n_{B})$ along with the equations for the evolution of the conformation tensors of the long $(\bm{A})$ and short worms $(\bm{B})$. According to this model, the equations for the variation of $n_{A}$, $n_{B}$, $\bm{A}$ and $\bm{B}$ are given in their non-dimensional forms as follows~\citep{vasquez2007network}:
\begin{equation}
    \label{nA}
    \mu\frac{Dn_{A}}{Dt} - 2\delta_{A} \nabla^{2}n_{A}  = \frac{1}{2} c_{B} n_{B}^{2} - c_{A}n_{A}
\end{equation}
\begin{equation}
    \label{nB}
    \mu\frac{Dn_{B}}{Dt} - 2\delta_{B} \nabla^{2}n_{B}  = - c_{B} n_{B}^{2} + 2 c_{A}n_{A}
\end{equation}
\begin{equation}
    \label{A}
    \mu \bm{A}_{(1)} + A -n_{A} \bm{I} -\delta_{A} \nabla^{2}\bm{A} = c_{B} n_{B} \bm{B} - c_{A} \bm{A}
\end{equation}
\begin{equation}
    \label{B}
    \epsilon \mu \bm{B}_{(1)} + B -\frac{n_{B}}{2} \bm{I} -\epsilon\delta_{B} \nabla^{2}\bm{B} = -2\epsilon c_{B} n_{B} \bm{B} + 2 \epsilon c_{A} \bm{A}
\end{equation}
Here the subscript $( )_{(1)}$ denotes the upper-convected derivative which is given as $\frac{\partial()}{\partial t} + \bm{U}\cdot \nabla () - \left( (\nabla \bm{U})^{T} \cdot () + ()\cdot \nabla \bm{U}\right)$. The non-dimensional parameters $\mu$, $\epsilon$ and $\delta_{A,B}$ are given as $\frac{\lambda_{A}}{\lambda_{eff}}$, $\frac{\lambda_{B}}{\lambda_{A}}$ and $\frac{\lambda_{A} D_{A,B}}{d^{2}}$ respectively where $\lambda_{B}$ is the relaxation time of the short chain $B$ and $D_{A, B}$ are the dimensional diffusivities of the long and short species A and B respectively. Furthermore, according to the VCM model, the non-dimensional breakage rate $(c_{A})$ of the long chain A into two equally sized small chains B depends on the local state of the stress field, and it is given by the expression as $c_{A} = c_{Aeq} + \mu \frac{\xi}{3}\left( \dot{\bm{\gamma}}: \frac{\bm{A}}{n_{A}} \right)$ whereas the reforming rate of the long chain A from the two short chains B is assumed to be constant which is given by the equilibrium reforming rate, i.e., $c_{B} = c_{Beq}$. Here the non-linear parameter $\xi$ is the scission energy required to break a long micelle chain into two shorter chains. The significance of this parameter is that as its value increases, the amount of stress needed to break the chain increases. 

\subsection{Single-species Giesekus constitutive equation}
In the single-species constitutive equation, the number density of the wormlike micelles remains constant due to the absence of breakage and reformation, and hence one doesn't need to solve any species conservation equation, as solved in the two-species VCM model. However, one has to solve the equation for the evaluation of the polymer conformation tensor (which is related to the stresses, as mentioned above) as follows
\begin{equation}
    \bm{A}_{(1)} + \bm{A} - \bm{I} = -\alpha \left( \bm{A} - \bm{I} \right) \cdot \left( \bm{A} - \bm{I} \right)  
\end{equation}
The dimensionless parameter $\alpha$ is known as the Giesekus mobility factor, and for $0 < \alpha <1$, the above equation is known as the Giesekus constitutive equation. This equation is derived based on the kinetic theory of closely packed polymer chains, and the mobility factor $\alpha$ is introduced in this model in order to take into account the anisotropic hydrodynamic drag on the polymer molecules~\citep{giesekus1982simple}.

\section{\label{NumDet}Numerical details}
All the governing equations, namely, mass, momentum, Giesekus, and VCM constitutive equations, have been solved using the finite volume method based open-source computational fluid dynamics code OpenFOAM~\citep{weller1998tensorial}. In particular, the recently developed rheoFoam solver available in rheoTool~\citep{rheoTool} has been used in the present study. A detailed discussion of the present numerical set up and its validation has been presented in our recent studies~\citep{sasmal2020flow,khan2020effect}, and hence it is not repeated here. The following boundary conditions were employed in order to solve the present problem. On the sphere surface, the standard no-slip boundary condition for the velocity, i.e., $U_{x} = U_{y} = 0$, is imposed whereas a no-flux boundary condition is assumed for both the stress and micellar number density, i.e., $\bm{n} \cdot \nabla \bm{A} = \bm{n} \cdot \nabla \bm{B} = 0$ and $\bm{n} \cdot \nabla n_{A} = \bm{n} \cdot \nabla n_{B} = 0$. It should be mentioned here that micelles may undergo a slip flow at the sphere surface, particularly if the sphere surface is roughened in nature~\citep{HadiBoundaryConditions}. However, in the present study, the sphere is assumed to be solid with a smooth surface, and hence the application of the no-slip boundary condition is justified at this stage. On the tube wall, $U_{x} = U$ and $U_{y} = 0$, and again no-flux boundary conditions for the stress and micellar number density are imposed. At the tube outlet, a Neumann type of boundary condition is applied for all variables except for the pressure for which a zero value is assigned here. A uniform velocity of $U_{x} = U$, a zero gradient for the pressure, and a fixed value for the micellar number density are employed at the tube inlet. Furthermore, the whole computational domain was sub-divided into 8 blocks in order to mesh it, as shown in figure~\ref{fig:Figure1}(b). Three different meshes of hexagonal block-structured, namely, M1 , M2,  and M3  with different numbers of cells on the sphere surface $(N_{s})$ as well as in the whole computational domain $(N_{t})$, were created for each blockage ratio. A schematic of three different mesh densities is shown in figure~\ref{fig:Figure1} for $BR = 0.33$.  In creating any mesh density, the cells were further compressed towards the sphere surface in order to capture the steep gradients of velocity, stress, and micellar concentration. After performing the standard mesh independent study, the mesh M2 (with $N_{s} = 240$ and $N_{t} = 74200$ for $BR = 0.33$ and $N_{s} = 240$ and $N_{t} = 78600$ for $BR = 0.1$) was found to be adequate to capture the flow physics for the whole range of conditions encompassed here for both the blockage ratios. Similarly, a time step size of $\Delta t = 0.000055$ was found to be suitable to carry out the present study. Finally, the two-dimensional and axisymmetric problem is realized in OpenFOAM by applying the standard wedge boundary condition (with wedge angle less than $5^0$) on the front and back surfaces of the computational domain, as schematically shown in figure~\ref{fig:Figure1}(c). The computational domain is kept one cell thick in the $\theta$ direction, and the axis of the wedge lies on the x-coordinate, as per the requirement for applying the wedge boundary condition~\citep{Openfoam}. Such simplification does not compromise the accuracy of the results as long as the flow is two-dimensional and axisymmetric, and also drastically reduces the computational cost compared to a full three-dimensional simulation.

\section{\label{Ressult}Results and discussion}
\begin{figure}
    \centering
    \includegraphics[trim=0cm 0cm 1cm 1cm,clip,width=12cm]{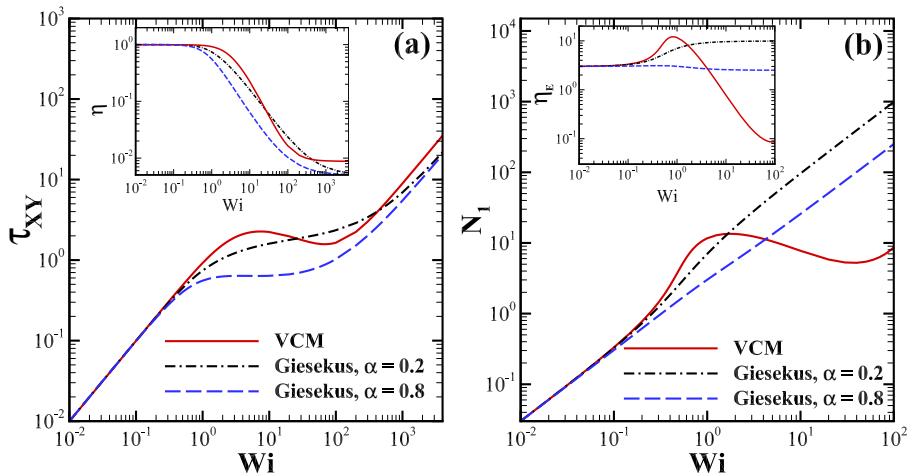}
    \caption{Variations of the non-dimensional shear stress with the shear rate (a) and non-dimensional first normal stress difference with the extension rate (b) in homogeneous shear and extensional flows respectively. Here the inset figures show the corresponding variations in the shear and extensional viscosities.}
    \label{fig:Figure2}
\end{figure}

The VCM model parameters chosen in the present study are as $\mu = 5.7$, $\epsilon = 4.5 \times 10^{-4}$, $\beta_{VCM} = 6.8 \times 10^{-5}$, $\xi = 0.7$, $n_{B}^{0} = 1.13$, $\delta_{A} = \delta_{B} = \delta = 10^{-3}$. These values are obtained by fitting the experimental results on small amplitude oscillatory shear (SAOS) and step strain experiments for a mixture of CPCl/NaSal added to water~\citep{pipe2010wormlike,zhou2014wormlike}. The rheological characteristics of the WLM solution with these parameter values in homogeneous shear and uniaxial extensional flows are shown in figure~\ref{fig:Figure2}. One can clearly see the two typical properties of a WLM solution, namely, the shear-thinning in shear flows and the extensional hardening and subsequent thinning in extensional flows in this figure. Additionally, the present WLM solution also shows a shear-banding phenomena. The corresponding parameters for the single-species Giesekus model are chosen as $\beta_{G} = 4.98 \times 10^{-3}$ and $\alpha = 0.2$ and 0.8. For the single-species model, at $\alpha=0.8$, the solution shows the shear banding and extensional thinning properties.  
Simulations were carried out for the shear Weissenberg number $(Wi_{S})$ of up to 2 for both the two-species VCM and single-species Giesekus models in the creeping flow regime. 
\begin{figure}
    \centering
   \includegraphics[trim=0cm 0cm 0cm 0cm,clip,width=13cm]{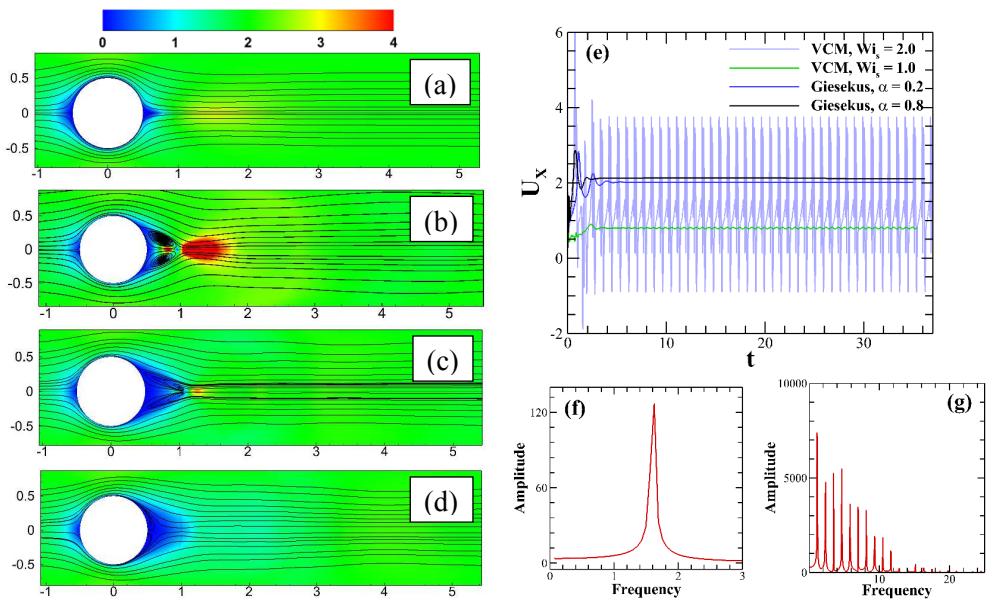}
    \caption{Representative streamlines and velocity magnitude plots for the Giesekus model (a) and for the VCM model at three different times, namely, $t = 50.7$ (b), $t = 50.8$ (c), and $t = 50.9$ at $Wi_{S} = 2.0$. Temporal variation of the stream-wise velocity component $(U_{X})$ for both Giesekus and VCM models (e). Power spectrum plot of the velocity fluctuations obtained with the VCM model at $Wi_{S} = 1.0$ (f) and $Wi_{S} = 2.0$ (g).}
    \label{fig:Stream}
\end{figure}
Up to the shear Weissenberg number of 0.6 (not shown here), the streamlines are attached to the sphere surface, and they follow a nice and order path without crossing to each other for both the single-species Giesekus and two-species VCM models. Hence there is a perfect fore and aft symmetry present in the streamline patterns, and also the flow remains steady up to this value of the Weissenberg number. 

However, as the Weissenberg number gradually starts to increase, clear differences have been observed in the flow patterns obtained with the Giesekus and VCM models. For instance, at $Wi_{\text{S}} = 2.0$ (sub figure~\ref{fig:Stream}(a)), the streamlines are still attached to the sphere surface for the single-species Giesekus model,  thereby suggesting no boundary layer separation happens for this model. Furthermore, the flow remains in the steady-state at this value of the Weissenberg number. This is confirmed by plotting the temporal variation of the stream-wise velocity at a probe location downstream of the sphere $(X = 1.0, Y = 0)$, sub figure~\ref{fig:Stream}(e). One can see that the velocity reaches a steady value with time. On the other hand, at the same Weissenberg number, for the two-species VCM model, separation of the boundary layer happens, and as a result, a small recirculation region is seen to form downstream of the sphere, sub figure~\ref{fig:Stream}(b). As time is further progressed, the wake detaches from the sphere surface, and its size becomes small (sub figure~\ref{fig:Stream}(c)), and ultimately, it is disappeared as can be seen in sub figure~\ref{fig:Stream}(d). This formation and disappearance of the wake is seen to repeat with time. It should be mentioned here that a weak recirculation region was seen besides the sphere but not downstream of the sphere for the falling sphere problem~\citep{chen2004flow}. All these observations suggest that the flow becomes unsteady at this Weissenberg number. This is further confirmed by plotting the stream-wise velocity in sub figure~\ref{fig:Stream}(e) at the same probe location as that obtained for the Giesekus model, and one can clearly see how the velocity fluctuates with time. In fact, the unsteadiness in the flow field appears at a much lower Weissenberg number of about $Wi_{S} = 0.6$ for the VCM model. Furthermore, a region of very high velocity magnitude is seen to appear at about one sphere diameter away downstream of the sphere at $t = 50.7$ (sub figure~\ref{fig:Stream}(b)). This indicates the presence of a \enquote{negative wake} downstream of the sphere. As time gradually increases, the magnitude of this region progressively decreases, and it is further shifted towards the downstream of the sphere, sub figure~\ref{fig:Stream}(b). Finally, it is vanished at a time $t = 50.9$, sub figure~\ref{fig:Stream}(c). On further increasing the time, it appears again, and these processes of appearance and disappearance of the negative wake downstream of the sphere repeat with time. This was also observed in the experiments for the falling sphere problem by Chen and Rothstein~\citep{chen2004flow}.

To further analyze the nature of the unsteadiness in the flow field, the power spectrum of these velocity fluctuations is plotted in sub figures~\ref{fig:Stream}(f) and (g) at shear Weissenberg numbers 1.0 and 2.0 respectively for the VCM model. At $Wi_{S} = 1.0$, the velocity fluctuations are characterized by a single and large narrow peak in the frequency spectrum, thereby suggesting the flow field to be a periodic one. On the other hand, at $Wi_{S} = 2.0$, the frequency spectrum of the velocity fluctuations is characterized by a broad range with the significant contribution from higher-order frequencies, sub figure~\ref{fig:Stream}(g). This suggests that the flow becomes quasi-periodic at this value of the Weissenberg number. Therefore, it can be seen that as the shear Weissenberg number gradually increases, the flow transits from a steady to periodic and then periodic to quasi-periodic for the two-species VCM model. This transition in the flow regime with the shear Weissenberg number is entirely in line with that observed experimentally for the falling sphere problem by Zhang and Muller~\citep{zhang2018unsteady}. Their experimental observations further found an irregular flow pattern after the quasi-periodic flow regime on further increasing the Weissenberg number. However, due to the high Weissenberg number numerical stability problem, we were unable to run simulations at Weissenberg numbers beyond 2, and hence this irregular flow pattern was not observed in our simulations within the ranges of conditions encompassed in this study. 

\begin{figure}
    \centering
   \includegraphics[trim=0cm 0.1cm 0cm 0cm,clip,width=13cm]{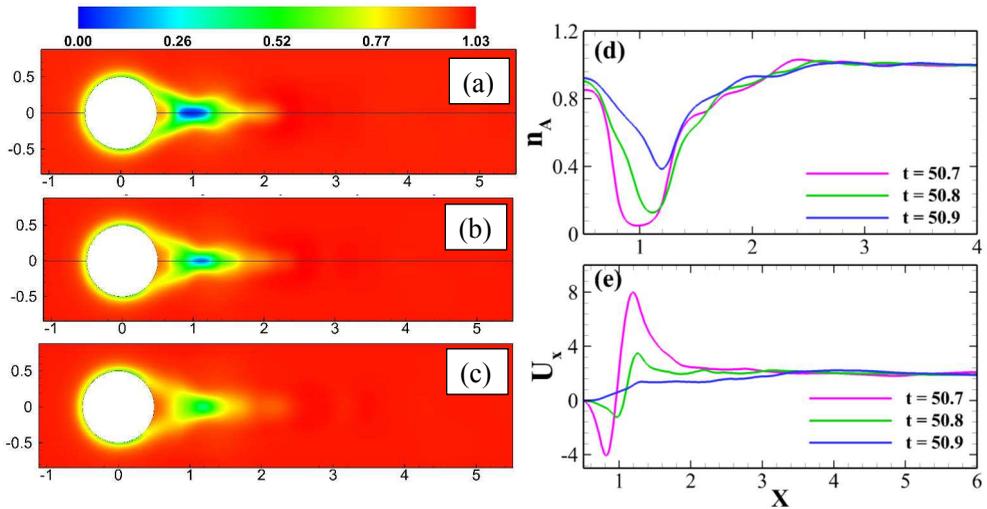}
    \caption{Surface plot of the long micelles number density at $Wi_{S} = 2.0$ and at three different times, namely, 50.7 (a), 50.8 (b) and 50.9 (c). Variation of the number density of long micelles (d) and stream-wise velocity component along the downstream axis of the sphere (e).}
    \label{fig:MicelleCon}
\end{figure}

Next, we explore the reason behind this unsteady motion, which is observed for the two-species VCM model but not seen for the single-species Giesekus model over the ranges of conditions considered in this study. The explanation goes as follows: at high values of the Weissenberg numbers, for instance, at $Wi_{S} = 2.0$ and $t = 50.7$, the long micelles tend to break into smaller ones downstream of the sphere due to the presence of high flow strength in this region, for instance, see sub figure~\ref{fig:MicelleCon}(a) where the surface plot of long micelles number density is presented at the same three different times as that used for showing the streamline and velocity magnitude plots in figure~\ref{fig:Stream}. To present it more quantitatively, the variation of long micelles number density along the downstream axis is plotted in sub figure~\ref{fig:MicelleCon}(d). It can be seen that the number density of long micelles shows a minimum at about $X = 0.5d$. The corresponding variation of the stream-wise velocity component along the downstream axis of the sphere is depicted in sub figure~\ref{fig:MicelleCon}(e). Due to the breakage of long micelles, the extensional load, which was carried out by the long micelles is unable to be carried out by the small micelles. This results in the formation of a small recirculation region downstream of the sphere. Due to the formation of this recirculation region, the velocity first gradually decreases, attains a minimum, and then gradually starts to increase. It then shows a peak and ultimately reaches the velocity far away downstream of the sphere. This stage of the unsteady motion was termed as the acceleration stage in case of the falling sphere problem~\citep{chen2004flow, mohammadigoushki2016sedimentation}. During this stage, a negative wake was formed, as shown in sub figure~\ref{fig:Stream}(b). The presence of a velocity overshoot in sub figure~\ref{fig:MicelleCon}(e) further confirms the existence of this negative wake downstream of the sphere. The region of velocity overshoot is seen to present just beside the region where the concentration of long micelles is minimum downstream of the sphere. As time is further progressed, new long micelles come into the downstream region, and the extensional stresses again start to develop. The velocity downstream of the sphere gradually starts to decrease, and the position at which the velocity changes its sign, further moves downstream, as can be seen in sub figure~\ref{fig:MicelleCon}(e) at $t = 50.8$. Ultimately, at $t = 50.9$, the velocity gradually reaches the far away downstream velocity without showing any minimum or peak in its profile. This means that at this time, no recirculation region as well as no negative wake, are present in the flow field (which are also evident in the streamline and velocity magnitude plots shown in figure~\ref{fig:Stream}), and under these conditions, the fluid behaves like a high elastic Boger fluid. This was called the deceleration stage in the unsteady motion for the falling sphere problem~\citep{chen2004flow, mohammadigoushki2016sedimentation}. Due to the deceleration in the flow field at later times, the breakage of long micelles also  decreases downstream of the sphere, as can be seen in sub figure~\ref{fig:MicelleCon}(e). On further increasing the time, the acceleration stage again appears, and it repeats with time. Therefore, this study provides evidence that the acceleration and deceleration motion past a steadily translating sphere in wormlike micellar solutions is solely caused due to the breakage of long micelles downstream of the sphere. This is further confirmed by the fact that the single-species Giesekus model shows a steady flow field under otherwise identical conditions. This explanation presented herein was also proposed by Chen and Rothstein~\citep{chen2004flow} in their experimental investigation for the unsteady motion of the falling sphere problem in WLM solutions. The present study further confirms their hypothesis about the onset of this instability using the two-species VCM model.   

\begin{figure}
    \centering
    \includegraphics[trim=1.5cm 0.5cm 0cm 0cm,clip,width=14cm]{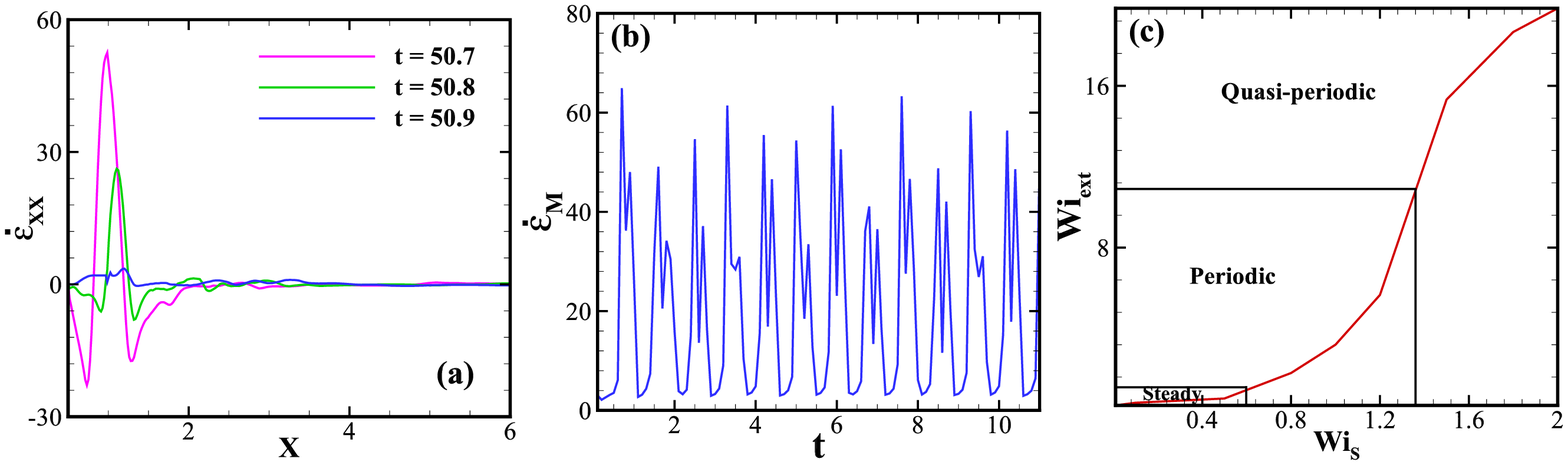}
    \caption{(a) Variation of the extension rate along the downstream axis of sphere at three different times and at $Wi_{\text{S}} = 2.0$ (b) Temporal variation of the maximum extension rate downstream of the sphere at $Wi_{\text{s}} = 2.0$ (e) Variation of the extensional Weissenberg number $(Wi_{\text{Ext}})$  versus the shear Weissenberg number$(Wi_{\text{s}})$.}
    \label{fig:Strain}
\end{figure}

Some studies associated with the falling sphere problem revealed that the onset of the sphere's unsteady motion is directly associated with the strong extensional flow behaviour downstream of the sphere, which ultimately leads to the breakage of long micelles~\citep{mohammadigoushki2016sedimentation,zhang2018unsteady}. Therefore,  we also calculate the extension rate along the downstream axis of the steadily translating sphere, defined as $\dot{\epsilon}_{XX} = \frac{\partial U_{X}}{\partial X}$. This is presented in sub figure~\ref{fig:Strain}(a) again at three different times, namely, $t = 50.7, 50.8$ and 50.9 and at $Wi_{S} = 2.0$. One can see that there is a large temporal variation present in the extension rate downstream of the sphere up to a distance of around $X = 1d$ from the rear stagnation point of the sphere, and beyond that, it becomes almost zero. At $t$ = 50.7, the variation in the strain rate is seen to be higher as compared to that seen at later times. At this time, the maximum breakage of long micelles occurs downstream of the sphere, as shown in sub figure~\ref{fig:MicelleCon}(a). The temporal variation of the maximum strain rate $(\dot{\epsilon}_{M})$ along the downstream axis of the sphere is shown in sub figure~\ref{fig:Strain}(b). It can be clearly seen that the maximum strain rate is varied quasi-periodically, and a large variation in its value is present. This is again, at least, qualitatively in line with that observed experimentally for the falling sphere problem~\citep{mohammadigoushki2016sedimentation,zhang2018unsteady}.
Likewise the experimental investigations on the falling sphere problem, we also define an extensional Weissenberg number based on the time-averaged values of this maximum strain rate as $Wi_{\text{Ext}} = \lambda_{eff} \dot{\epsilon}_{M}$. The variation of this extensional Weissenberg number with the corresponding shear Weissenberg number is shown in sub figure~\ref{fig:Strain}(c). From this figure, it is seen that the value of the extensional Weissenberg increases with the shear Weissenberg number, and the transition from steady to unsteady periodic is marked by a sharp increase in the value of the extensional Weissenberg number, which is again, in line with the corresponding experimental observations for the falling sphere problem~\citep{mohammadigoushki2016sedimentation,zhang2018unsteady}.

\begin{figure}
    \centering
    \includegraphics[trim=0cm 0.1cm 0cm 0cm,clip,width=13cm]{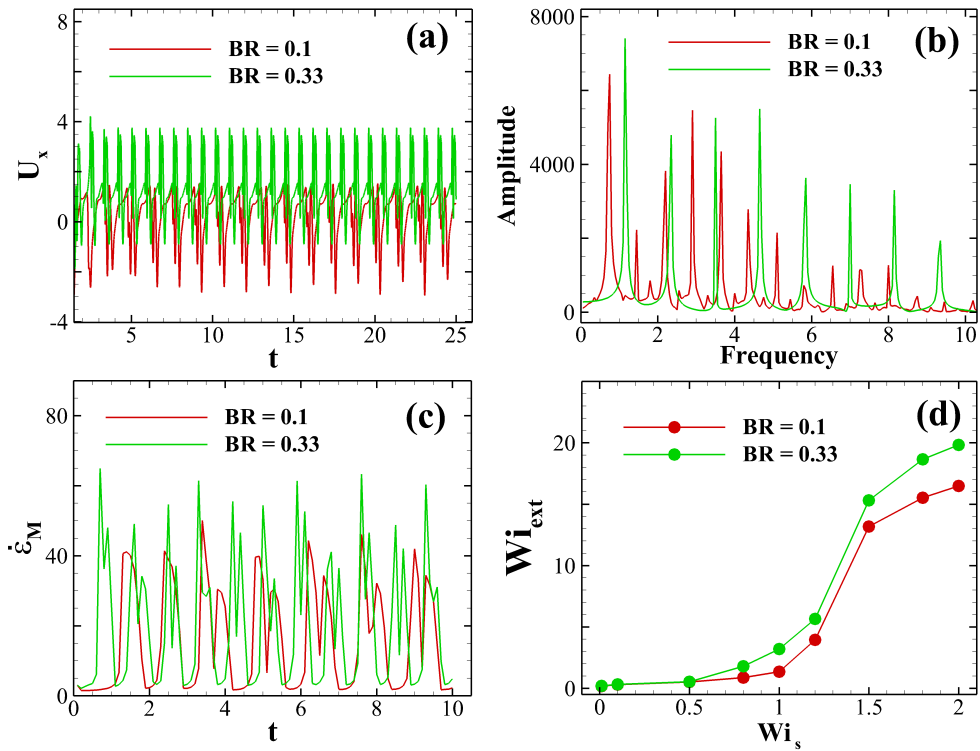}
    \caption{Effect of the sphere to tube diameter ratio or the blockage ratio (BR) on the temporal variation of the stream-wise velocity (a), power spectrum plot (b) and the temporal variation of the maximum extension rate (c) downstream of the sphere at $Wi_{s} = 2.0$. Effect of the blockage ratio on the variation of the extensional Weissenberg number with the shear Weissenberg number (d).}
    \label{fig:EffectofBR}
\end{figure}

Finally, the effect of the sphere to tube diameter ratio on the onset and generation of this unsteady motion past the sphere is discussed. Simulations were carried out at another value of $\frac{d}{D} = 0.1$ in order to compare with flow characteristics as discussed above at $\frac{d}{D} = 0.33$ under otherwise identical conditions. Figure~\ref{fig:EffectofBR}(a) shows the temporal variation of the stream-wise velocity at two sphere to tube diameter ratios at a probe location (X = 1, Y = 0) downstream of the sphere and at a shear Weissenberg number of $Wi_{s} = 2.0$. One can see that the velocity field again shows a similar quasi-periodic nature at $\frac{d}{D}$ = 0.1 as that seen at $\frac{d}{D}$ = 0.33. However, the magnitude of the velocity during the acceleration stage decreases, whereas it increases during the deceleration stage. Moreover, the magnitude of the velocity fluctuations slightly decreases with the decreasing values of the sphere to the tube diameter ratio. This is clearly evident in the power spectrum plot (sub figure~\ref{fig:EffectofBR}(b)) wherein the amplitude of the maximum frequency spectrum is seen to be slightly higher at $\frac{d}{D} = 0.33$ than that seen at $\frac{d}{D} = 0.1$. At $\frac{d}{D} = 0.1$, a similar transition in the velocity field is seen as that observed at $\frac{d}{D} = 0.33$, i.e., it transits from steady to unsteady periodic to unsteady quasi-periodic upon increasing the Weissenberg number. However, the onset of the unsteady motion is slightly delayed as the sphere to tube diameter ratio decreases. For instance, at $\frac{d}{D} = 0.33$, the unsteady motion starts at $Wi_{s} = 0.6$ whereas it starts at around $Wi_{s} = 1.0$ for the
sphere to tube diameter ratio 0.1. The reason behind this can be explained as follows: as the sphere to tube diameter ratio decreases, the extensional flow strength downstream of the sphere also decreases due to the presence of less confinement. This can be seen in sub sub figure~\ref{fig:EffectofBR}(c) wherein the temporal variation of the maximum extension rate downstream of the sphere is seen to be less at $\frac{d}{D} = 0.1$ in comparison to that seen at $\frac{d}{D} = 0.33$. As a result, the time-averaged extensional Weissenberg number is also found to be less for the former one in comparison to the latter one  under otherwise identical conditions, sub figure~\ref{fig:EffectofBR}(d). This lowering in the extensional flow strength downstream of the sphere tends to decrease the breakage of the micelles in this region, which in turn, delays the tendency of appearing the unsteady motion. This also further confirms the hypothesis that the unsteady motion past a sphere translating steadily in micellar solutions is caused due to the breakage of stretched micelles downstream of the sphere.

\section{\label{Con}Conclusions}
 This study presents an extensive numerical investigation of the flow characteristics past a sphere translating steadily along the axis of a cylindrical tube filled with wormlike micellar solutions. For doing so, the present study uses the two-species VCM (Vasquez-Cook-McKinley) and single-species Giesekus constitutive models for representing the rheological behaviour of wormlike micellar solutions. Over the ranges of conditions encompassed in this study,  a transition of the flow field downstream of the sphere from a steady to unsteady periodic to unsteady quasi-periodic regime, is seen as the shear Weissenberg number gradually increases. A similar transition in the velocity field was also observed in the experiments on the sedimentation of a sphere in wormlike micellar solutions. The onset of this unsteady motion is marked by a steep increase in the value of the extensional Weissenberg number, defined downstream of the sphere based on the maximum extension rate, once again in accordance with the experiments on the falling sphere problem. Due to this increase in the extensional flow strength downstream of the sphere, the breakage of long micelles occurs, which thereby causing the unsteady motion in the flow field downstream of the sphere. This is further confirmed as the single-species Giesekus model for the wormlike micellar solution predicts a steady velocity field under otherwise identical conditions. This explanation is in line with that proposed by the earlier experimental investigations carried out in the literature for the sedimentation of a sphere in wormlike micellar solutions. Furthermore, it is seen that the onset of this unsteady motion is delayed as the sphere to tube diameter ratio decreases due to the decrease in the extensional flow strength downstream of the sphere.
 
 Although a very good qualitative agreement is found between the present numerical predictions on the steadily translating sphere and the experimental findings on the sedimentation of sphere; however, it should be mentioned here that the present simulation set up does not mimic the exact experimental settings for the falling sphere problem. In the experiments on the falling sphere problem, the sphere may rotate or undergo lateral motion or even may not reach a terminal velocity. Therefore, to realize the experimental settings on the falling sphere problem, the governing equations, namely, continuity, momentum and micelle constitutive equations need to be solved in full three-dimensional numerical settings along with an equation for the sphere motion. In contrast to this, in the present study, the sphere is assumed to be translating steadily along the axis of the tube, and the problem is solved in a co-ordinate system which is centered on and travelling with the sphere. This could be a situation in the corresponding experiments on the falling sphere problem when it reaches to its terminal velocity without rotation and lateral motion. Although this is hardly a case in actual experiments; however, with this simplified problem, for the first time, we have provided the eveidence that the unsteady motion past a sphere is caused due to the breakage of long micelles, resulting from an increase in the extensional flow strength downstream of it. We believe that the analysis shown in this study will further support the hypothesis presented earlier for the unsteady motion of a falling sphere in micellar solutions. In our future study, we plan to carry out full three-dimensional numerical simulations with the exact settings as that followed in the experiments for the sedimentation of sphere in micellar solutions. This will give us the opportunity to conduct a more accurate and quantitative comparison with the experiments carried out for the falling sphere problem.  
 
 \section{Acknowledgements}
Author would like to thank IIT Ropar for providing funding through the ISIRD research grant (Establishment1/2018/IITRPR/921) to carry out this work. Author would also like to thank Mr. Anant Chauhan for his help in the meshing of the geometry in OpenFOAM. 

\bibliographystyle{jfm}
\bibliography{jfm-instructions}

\end{document}